\documentclass[preprintnumbers,prd,superscriptaddress,preprint]{revtex4-1}

\usepackage{CJK}
\usepackage{amsmath}
\usepackage{subfigure}
\usepackage{enumerate}
\usepackage{amssymb}
\usepackage{amsfonts}
\usepackage{mathrsfs}
\usepackage{tensor}
\usepackage{latexsym}
\usepackage{bm}
\usepackage{amscd}
\usepackage{physics} 
\usepackage{graphicx}
\usepackage{epsfig}
\usepackage{hhline,multirow}
\usepackage{dcolumn}
\usepackage{url}
\usepackage[colorlinks=true,linkcolor=blue,citecolor=blue,urlcolor=blue]{hyperref}
\usepackage{color}
\usepackage{xcolor}
\usepackage{indentfirst}
\usepackage{subfigure}
\usepackage{psfrag}
\usepackage{slashed}
\usepackage{float}
\usepackage{setspace}
\usepackage{mathtools}
\allowdisplaybreaks[2]

\begin{document}
\title{Translation gauge field theory of gravity in Minkowski spacetime}
\author{Hang Li}
\affiliation{Institute of High Energy Physics, CAS, P. O. Box
	918(4), Beijing 100049, China}
\affiliation{School of Physical Sciences, University of Chinese Academy of Sciences, Beijing 101408, China}

\author{P. Wang}
\affiliation{Institute of High Energy Physics, CAS, P. O. Box
	918(4), Beijing 100049, China}
\affiliation{School of Physical Sciences, University of Chinese Academy of Sciences, Beijing 101408, China}

\begin{abstract}
The gravitational field $h_{\mu\nu}$ with spin-2 is introduced naturally by the requirement that the Lagrangian is locally translation invariant in Minkowski spacetime. The interactions between the $h_{\mu\nu}$ and spin-$\frac12$, 0, 1 matter fields are obtained along with the Lagrangian for the gravitational field including self-interactions. The deflection angle of light when it passes through the sun is calculated with different gauge conditions as an example. Our leading-order result is the same as that from general relativity, although the basic ideas are different. It is interesting that gravity can be described in a similar way to other fundamental interactions in Minkowski spacetime, and it may provide a new scenario for the Universe.

\end{abstract}

\maketitle

\section{Introduction}
Einstein's general relativity takes gravity as a geometric property of spacetime and has successfully described a variety of gravitational phenomena at several scales \cite{mtw,will}. Many experiments have verified that gravitation is a phenomenon of curved spacetime, i.e., the underlying gravitational theory should be a metric one. The great success of general relativity, however, has not stopped alternatives from being propounded. There are many modified gravity theories, such as scalar-tensor theory \cite{Bergmann,Wagoner}, vector-tensor theory \cite{Jacobson}, bimetric theory \cite{Rosen,Drum}, tensor-vector-scalar theory \cite{Bekenstein1,Bekenstein2}, $f(R)$ theory \cite{Nojiri,Sotiriou}, Ho\v{r}ava-Lifschitz gravity \cite{Horava,Padilla}, Galileon theory \cite{Nicolis}, and models of extra dimensions, including Kaluza-Klein \cite{Nordstrom}, Randall-Sundrum \cite{Randall1,Randall2}, and Dvali-Gabadadze-Porrati gravity \cite{Dvali}. The parameterised post-Newtonian (PPN) formalism was proposed to compare and assess various gravity theories \cite{will}. For a theoretical review, see, for example, Ref.~\cite{modify}.

In contrast with gravity, the electromagnetic, weak, and strong interactions are all described as gauge theories, which are related to some internal symmetries. Einstein himself attempted to unify the electromagnetic and gravitational fields by introducing a tetrad or vierbein field \cite{Einstein,vierbein,Santos1}. This is known as the teleparallel theory of gravity. Later, M\"{o}ller revived teleparallel theory when he introduced his energy-momentum complex to solve the localization of energy and momentum in general relativity \cite{moller}. After this, Hayashi et al. proposed the new general relativity as a teleparallel theory described by the Weitzenb\"{o}ck connection obtained due to the condition of absolute parallelism \cite{hayashi0,hayashi1,hayashi2,hayashi3,terg2}. This implies that the new general relativity can be presented as a gauge theory of the translation group. Localizing the translation group will result in the Weitzenb\"{o}ck spacetime. In fact, after the emergence of gauge concept, several authors began to attempt to derive the gravitational interaction by gauging the Lorentz and Poincar\'{e} groups \cite{utiyama1,utiyama2,kibble}. Applying these external groups in Minkowski spacetime will lead to a gauge theory related to gravity \cite{Hehl}. Fixing the parameters emerging from the decomposition of the Weitzenb\"{o}ck torsion can render a new general relativity to the teleparallel equivalent of general relativity. In the teleparallel equivalent of general relativity, all the effects of gravity are encoded in the torsion tensor. Some authors established a more mathematically sound framework for new general relativity using Cartan geometry \cite{terg1,Huguet}. The energy–momentum distribution in the framework of this teleparallel theory of gravity has also been discussed \cite{e-m1,e-m2,e-m3}. Besides the different gravity theories obtained by gauging different symmetry groups \cite{Sard,Hehl2,Sauro}, there are several other gauge theories of gravity, such as nonlocal translation gauge theory \cite{mash} and $f(T)$ theories of modified teleparallel gravity \cite{Ferraro1,Ferraro2}.

In contrast with general relativity and the above gauge theories that use the geometrical gravity approach to explain gravitation, the ``field'' gravity theory was constructed similarly to other fundamental physical fields using the Lagrangian formalism of relativistic quantum field theory in Minkowski spacetime \cite{thirring,Pauli,Feynman}. Many authors have derived Einstein's field equations of general relativity from spin-2 field theory \cite{Gupta,Kraichnan,Deser,Arkani}. From the free spin-2 Lagrangian which is infinitesimally translation invariant, and the interaction between the gravitational field and matter fields, one can obtain the equation of motion for the spin-2 field. However, it is not self-consistent because in the gauge field theory of gravity, the spin-2 gravitational field itself should have the energy-momentum as well as other matter fields, which is also the source of gravity. Simply adding the energy-momentum tensor of the gravitational field to the tensor $T_{\mu\nu}$ of matter fields is still not consistent because the modified Lagrangian, which includes the self-interaction of gravitational field, will generate a new energy-momentum tensor, and this new energy-momentum tensor will again result in a new modified Lagrangian. This is an infinite process. An iterative gravity field theory in Minkowski spacetime was partly developed, where the theory was constructed step by step using an iteration procedure so that at each step, all physical properties of the energy-momentum tensor of the gravity field are under control \cite{Soko1,Soko2,Bary1,Bary4}. Each step of iteration is described by linear gauge-invariant field equations with fixed sources. It has been noted that an ostensible field theory of gravity in flat spacetime is actually general relativity. All the derivations of general relativity from spin-2 field theory are based on some additional assumptions equivalent to the geometrization of the gravitational interaction \cite{mtw,Strau1,Pad}. It is still an open question whether the field gravity theory is experimentally equivalent to the geometrical general relativity. Some tests that can clarify whether the gravity is the curvature of spacetime or a matter field in Minkowski spacetime, as is the case of other physical forces, have been reviewed \cite{Bary2,Bary3}. 

General relativity is constructed in curved Riemann spacetime, while the gauge theory of gravity usually deforms the underlying Minkowski spacetime and leads to new geometry. In these gauge theories, the gravitational field is represented by the metric tensor. The field theory of gravity applies the iteration process starting from the free spin-2 field. The interaction Lagrangian has been introduced as a principle of universality to replace the equivalence principle used in the geometrical approach \cite{Bary1}. However, in this study, the gravitational tensor field $h_{\mu\nu}$ is introduced naturally to guarantee the local translation invariance of the Lagrangian, and it has nothing to do with the metric. The spacetime is always flat, as in the field gravity theory. However, our translation gauge group in not infinitesimal. In addition, the principle of universality, i.e., the interaction between the gravitational and matter fields, is derived from gauge invariance rather than introduced as an assumption. In Sec.~\ref{sec-2}, the gravitational interactions between the tensor field $h_{\mu\nu}$ and matter field with spin-0, 1/2 and 1, and the self-interactions for $h_{\mu\nu}$ are all obtained with the same requirement that the Lagrangian is locally translation invariant. We discuss the deflection of light when it passes through the sun as an example in Sec.~\ref{sec-3}. Finally, Sec.~\ref{sec-4} presents a summary.

\section{Translation invariant Lagrangian}\label{sec-2}
The free Lagrangian $\mathcal{L}^0_{\text{F}}=\bar{\psi}(i\gamma^\mu \partial_\mu-m)\psi$ for a quark or lepton field is invariant under the global $U(1)$ transformation $\psi(x)\to e^{ie\alpha}\psi(x)$, where $\alpha$ is a constant. With the N\"{o}ther theory, the electromagnetic current $J^\mu(x)$ can be obtained as $J^\mu(x) = e\bar{\psi}(x) \gamma^\mu \psi(x)$. The interaction between the quark/lepton field and photon field $A_\mu(x)$ can then be expressed as $J^\mu(x) A_\mu(x)$. The photon field can be naturally introduced with the proper transformation property, and the interaction between the quark/lepton and photon fields can be automatically obtained if we assume that the total Lagrangian is locally $U(1)$ invariant, which means $\alpha$ is spacetime dependent. The standard model for strong and electroweak interactions is established in the similar way. 
One may wonder whether the gravitational interaction can also be constructed in this way. For gravity, the current is related to the energy-momentum tensor, which is generated from the translation invariance of the Lagrangian. In this section, we discuss the interactions between gravitational field and matter fields, including Fermion and Boson fields, and the self-interactions of gravitational field.

\subsection{Lagrangian for Fermion field}
Under the local translation, a Dirac field transforms as 
\begin{equation}\label{eq-psitr}
\psi(x)\rightarrow \psi'(x) = \psi(x'), 
\end{equation}
where $x'^\mu=x^\mu+\theta^\mu(x)$. The free Lagrangian $\mathcal{L}^0_{\text{F}}$ transforms as 
\begin{eqnarray}
\mathcal{L}^0_{\text{F}}(x)&\to&\frac{i}{2}\left[\bar{\psi}(x^{\prime})\gamma^\mu\partial_\mu\psi(x^{\prime})-(\partial_\mu\bar{\psi}(x^{\prime}))\gamma^\mu\psi(x^{\prime})\right]-m\bar{\psi}(x^{\prime})\psi(x^{\prime})\nonumber \\
&=&\frac{i}{2}\left[\bar{\psi}(x^{\prime})\gamma^\mu\partial'_\mu\psi(x^{\prime})-(\partial'_\mu\bar{\psi}(x^{\prime}))\gamma^\mu\psi(x^{\prime})\right]-m\bar{\psi}(x^{\prime})\psi(x^{\prime})\nonumber\\
&&+\frac{i}{2}\left[\bar{\psi}(x^{\prime})\gamma^\mu\partial'^\nu\psi(x^{\prime})-(\partial'^\nu\bar{\psi}(x^{\prime}))\gamma^\mu\psi(x^{\prime})\right]\partial_\mu\theta_\nu(x),
\end{eqnarray}
where the first line on the right hand side is the same as the original Lagrangian, except the argument $x$ is replaced by $x'$. To cancel the second line, the tensor field $h_{\mu\nu}(x)$ must be introduced to the Lagrangian as
$g\bar{\psi}(x)i\gamma^\mu h_{\mu\nu}(x) \partial^\nu\psi(x)
-g(h_{\mu\nu}(x)\partial^\nu\bar{\psi}(x))\gamma^\mu\psi(x)$
with the transformation property
\begin{equation}\label{eq-transh}
h_{\mu\nu}(x)\partial_\rho\to \left[h_{\mu\nu}(x')-\frac{1}{g}\partial_\mu\theta_\nu(x)\right]\partial'_\rho,
\end{equation}
where $g$ is the coupling constant. Therefore, the total Lagrangian $\mathcal{L}_{\text{F}}(x)$ is locally translation invariant, which is expressed as 
\begin{eqnarray}\label{eq-LDirac}
\mathcal{L}_{\text{F}}(x)&=&\frac{i}{2}(\eta_{\mu\nu}+gh_{\mu\nu}(x))\left[\bar{\psi}(x)\gamma^\mu\partial^\nu\psi(x)-(\partial^\nu\bar{\psi}(x))\gamma^\mu\psi(x)\right]-m\bar{\psi}(x)\psi(x)\nonumber \\
&=&\mathcal{L}^0_{\text{F}}(x)+gh_{\mu\nu}(x)\tilde{T}_{\text{F}}^{\mu\nu}(x)\nonumber \\
&=&\frac{i}{2}\left[\bar{\psi}(x)\gamma^\mu D_\mu\psi(x)-(D_\mu\bar{\psi}(x))\gamma^\mu\psi(x)\right]-m\bar{\psi}(x)\psi(x),
\end{eqnarray}
where $\tilde{T}_{\text{F}}^{\mu\nu}(x)$ is the asymmetric tensor for the Dirac field, expressed as
\begin{equation}
\tilde{T}_{\text{F}}^{\mu\nu}(x)=\frac{i}{2}\left[\bar{\psi}(x)\gamma^\mu\partial^\nu\psi(x)-(\partial^\nu\bar{\psi}(x))\gamma^\mu\psi(x)\right]. 
\end{equation}
$D_\mu$ is the covariant derivative
$D_\mu = \partial_\mu + gh_{\mu\nu}(x)\partial^\nu$, and $\eta_{\mu\nu}=\text{diag}\{+1,-1,-1,-1\}$ is the metric tensor of Minkowski spacetime. Because $\partial_\rho=(\delta_\rho^\sigma+\partial_\rho \theta^\sigma(x))\partial'_\sigma$, when $\partial_\mu \theta_\nu(x)$ is small, the transformation property
of $h_{\mu\nu}(x)$ can be obtained order by order. For example, at leading order, $h_{\mu\nu}(x)$ transforms as
\begin{equation}\label{eq-trans1}
h_{\mu\nu}(x)\to h_{\mu\nu}(x')-\frac{1}{g}\partial_\mu\theta_\nu(x).
\end{equation}
At next-to-leading order, it transforms as
\begin{equation}\label{eq-trans2}
h_{\mu\nu}(x)\to h_{\mu\nu}(x')-\frac{1}{g}\partial_\mu\theta_\nu(x) - \left[h_{\mu\rho}(x')
-\frac{1}{g}\partial_\mu\theta_\rho(x)\right]\partial^\rho\theta_\nu(x).
\end{equation}
From Eq.~(\ref{eq-LDirac}), we can see that the locally translation invariant Lagrangian is obtained from the free Lagrangian via the replacement of $\partial_\mu$ with $D_\mu$. It is equivalent to replace $\eta_{\mu\nu}$ with $\eta_{\mu\nu} + gh_{\mu\nu}(x)$. It is interesting that the tensor $\tilde{T}_{\text{F}}^{\mu\nu}(x)$ is not exactly the same as the energy-momentum tensor obtained by the N\"{o}ther theory from the global translation symmetry. This is different from the electromagnetic case, where the electromagnetic current in the interaction $A_\mu(x)J^\mu(x)$ obtained from the local $U(1)$ symmetry is the same as that obtained from the global $U(1)$ symmetry. Owing to the derivative in the tensor current $\tilde{T}_{\text{F}}^{\mu\nu}(x)$, it is not invariant under the local translation, i.e. ,
\begin{equation}
\tilde{T}_{\text{F}}^{\mu\nu}(x)\rightarrow \tilde{T}_{\text{F}}^{\mu\rho}(x')(\delta_\rho^\nu + \partial^\nu \theta_\rho(x)). 
\end{equation}
This is similar to the color current in the QCD case, where the current is not invariant under the local $SU(3)_c$ transformation. However, because the derivative in the  tensor current transforms together with the gravitational field $h_{\mu\nu}$ according to Eq.~(\ref{eq-transh}), the interaction $h_{\mu\nu}(x)\tilde{T}_{\text{F}}^{\mu\nu}(x)$ transforms as 
\begin{equation}
h_{\mu\nu}(x)\tilde{T}_{\text{F}}^{\mu\nu}(x) \rightarrow \left[h_{\mu\nu}(x')-\frac{1}{g}\partial_\mu\theta_\nu(x)\right]\tilde{T}_{\text{F}}^{\mu\nu}(x').
\end{equation}
This is comparable to the $U(1)$ transformation of the electromagnetic interaction $A_\mu(x)J^\mu(x)$ 
\begin{equation}
A_\mu(x)J^\mu(x) \rightarrow \left[A_\mu(x) - \frac{1}{e}\partial_\mu \theta(x)\right] J^\mu(x).
\end{equation}

The tensor $h_{\mu\nu}(x)$ describes the gravitational field of spin-2 with gauge freedoms owing to the invariance of translation. It is convenient to set $h_{\mu\nu}(x)$ as symmetric, traceless, and divergence free \cite{Adrian}, i.e.,
\begin{equation}\label{eq-gauge}
h_{\mu\nu}(x)=h_{\nu\mu}(x),\quad\eta^{\mu\nu}h_{\mu\nu}(x)=0,\quad\partial^\mu h_{\mu\nu}(x)=0.
\end{equation}
With the above constraints, the Lagrangian $\mathcal{L}_{\text{F}}(x)$ can be written as
\begin{equation}\label{eq-Lsy}
\mathcal{L}_{\text{F}}(x)=\mathcal{L}^0_{\text{F}}(x)+gh_{\mu\nu}(x)T_{\text{F}}^{\mu\nu}(x),
\end{equation}
where $T_{\text{F}}^{\mu\nu}(x)$ is the symmetric Belinfante--Rosenfeld energy--momentum tensor of the Dirac field, expressed as
\begin{equation}
T_{\text{F}}^{\mu\nu}(x)=\frac{i}{4}(\bar{\psi}\gamma^\mu\partial^\nu\psi+\bar{\psi}\gamma^\nu\partial^\mu\psi)-\frac{i}{4}((\partial^\mu\bar{\psi})\gamma^\nu\psi+(\partial^\nu\bar{\psi})\gamma^\mu\psi)-\eta^{\mu\nu}\mathcal{L}^0_{\text{F}}.
\end{equation}
Neither the asymmetric part of $\tilde{T}_{\text{F}}^{\mu\nu}(x)$ nor the term $\eta^{\mu\nu}\mathcal{L}^0_{\text{F}}$ has contributions to the gravitational interaction $gh_{\mu\nu}(x)T_{\text{F}}^{\mu\nu}(x)$ when $h_{\mu\nu}(x)$ is symmetric and traceless. Under the local translation, the Lagrangian of Eq.~(\ref{eq-Lsy}) will generate an additional term,
\begin{eqnarray}
\mathcal{L}_{\text{F}}(x) &\rightarrow& \mathcal{L}_{\text{F}}(x')+
\frac{i}{4}(\partial_\mu \theta_\nu(x) - \partial_\nu \theta_\mu(x))\left[\bar{\psi}(x^{\prime})\gamma^\mu\partial'^\nu\psi(x^{\prime})-(\partial'^\nu\bar{\psi}(x^{\prime}))\gamma^\mu\psi(x^{\prime})\right] \nonumber \\
&& + (\partial_\mu\theta^\mu(x))\left\{\frac{i}{2}\left[\bar{\psi}(x^{\prime})\gamma^\nu\partial'_\nu\psi(x^{\prime})-(\partial'_\nu\bar{\psi}(x^{\prime}))\gamma^\nu\psi(x^{\prime})\right]-m\bar{\psi}(x^{\prime})\psi(x^{\prime})\right\}.
\end{eqnarray}
In other words, the Lagrangian (\ref{eq-Lsy}) is only invariant under the symmetric and traceless translation with
\begin{equation}
\partial_\mu\theta_\nu(x)=\partial_\nu\theta_\mu(x),\quad\partial_\mu\theta^\mu(x)=0.
\end{equation}
We should mention that, in general, the gravitational field and tensor current of the matter field do not have to be symmetric. Eq.~(\ref{eq-gauge}) is the gauge condition rather than the equation of motion for the gravitational field $h_{\mu\nu}(x)$. This is just one specified choice. The translation invariance gives the field $h_{\mu\nu}(x)$ have gauge freedoms. 

\subsection{Lagrangian for Boson field}
For the spin-0 case, the free Lagrangian is written as
\begin{equation}
\mathcal{L}^0_{\text{S}}=\frac12\partial_\mu\phi\partial^\mu\phi-\frac12m^2\phi^2.
\end{equation}
In the above and most of the following equations, the argument $x$ is omitted for convenience.
Under the transformation of translation, $\phi(x)$ is changed to be $\phi(x+\theta(x))$.
By replacing $\partial_\mu$ with the covariant derivative $D_\mu$, the free Lagrangian $\mathcal{L}^0_{\text{S}}$ can be transformed into the locally translation invariant Lagrangian
\begin{equation}
\mathcal{L}_{\text{S}}=\frac12D_\mu\phi D^\mu\phi-\frac12m^2\phi^2 = \mathcal{L}_{\text{S}}^0+gh_{\mu\nu}\tilde{T}_{\text{S}}^{\mu\nu}+\frac12g^2h_{\mu\rho}h^{\mu\sigma}\partial^\rho\phi\partial_\sigma\phi,
\end{equation}
where $\tilde{T}_{\text{S}}^{\mu\nu}$ is the tensor current for the spin-0 field, expressed as 
\begin{equation}
\tilde{T}_{\text{S}}^{\mu\nu}=\partial^\mu\phi\partial^\nu\phi.
\end{equation}
Unlike the spin-$\frac12$ case, in the spin-0 case, the tensor current $\tilde{T}_{\text{S}}^{\mu\nu}$, which couples with the gravitational field $h_{\mu\nu}$, is symmetric without requirement that $h_{\mu\nu}$ is symmetric. In the spin-0 case, besides the leading order interaction $gh_{\mu\nu}\tilde{T}_{\text{S}}^{\mu\nu}$, there is a high order interaction $\frac12 g^2 h_{\mu\rho}h^{\mu\sigma}\partial^\rho \phi\partial_\sigma \phi$. Certainly, as in the spin-$\frac12$ case, we can also choose the symmetric and traceless gauge. The leading interaction term can then be expressed as $gh_{\mu\nu}T_{\text{S}}^{\mu\nu}$, where $T_{\text{S}}^{\mu\nu}$ is the canonical energy-momentum tensor
\begin{equation}
T_{\text{S}}^{\mu\nu}=\partial^\mu\phi\partial^\nu\phi - \eta^{\mu\nu}{\cal L}^0_{\text{S}}.
\end{equation}
It should be noted that for the matter fields with any spin, they transform in the manner of Eq.~(\ref{eq-psitr}), which is different from Refs.~\cite{Adrian,Santos2}, where the scalar, fermion, and vector fields have different transformation properties. As a result, there was no interaction between $h_{\mu\nu}$ and the scalar fields \cite{Santos2}. However, in our method, $D_\mu$ is the same for all matter fields with any spin.

For the massive spin-1 field, the naive free Lagrangian can be written as
\begin{equation}
\mathcal{L}^0_{\text{Naive}}=-\frac12\partial_\mu A_\nu\partial^\mu A^\nu+\frac{m^2}{2}A_\mu A^\mu.
\end{equation}
With the same approach, the locally translation invariant Lagrangian can be written as
\begin{equation}\label{eq-Lnvector}
\mathcal{L}_{\text{Naive}}=-\frac12D_\mu A_\nu D^\mu A^\nu+\frac{m^2}{2}A_\mu A^\mu=\mathcal{L}^0_{\text{Naive}}+gh_{\mu\nu}\tilde{T}_{\text{Naive}}^{\mu\nu}-\frac12 g^2h_{\mu\rho}h^{\mu\sigma}\partial^\rho A_\nu\partial_\sigma A^\nu,
\end{equation}
where $\tilde{T}_{\text{Naive}}^{\mu\nu}$ is the symmetric tensor current for the above naive Lagrangian, expressed as
\begin{equation}
\tilde{T}_{\text{Naive}}^{\mu\nu}=-\partial^\mu A^\rho\partial^\nu A_\rho.
\end{equation}
Again, besides the leading order interaction, the high order interaction appears. 
The vector $A_\mu(x)$ has four degrees of freedom, whereas the massive vector particle has three degrees of freedom. To describe the massive vector particle, the correct Lagrangian is the so-called Proca Lagrangian ${\cal L}_{P}$ constructed by the strength tensor $F_{\mu\nu}$ \cite{Schwartz}. In the massless limit, for example, for a photon field, the free Lagrangian is locally $U(1)$ gauge invariant and expressed as
\begin{equation}
{\cal L}_{\text{V}}^0 = -\frac14F_{\mu\nu}F^{\mu\nu},
\end{equation}
where
\begin{equation}
F_{\mu\nu}=\partial_\mu A_\nu-\partial_\nu A_\mu.
\end{equation}
The interaction between the photon and gravitational fields can be obtained via the requirement of translation invariance. The locally translation invariant Lagrangian for the photon field is written as
\begin{equation}
{\cal L}_{\text{V}} = -\frac14(D_\mu A_\nu-D_\nu A_\mu)(D^\mu A^\nu-D^\nu A^\mu).
\end{equation}
After the expansion of the above Lagrangian, we can get
\begin{equation}\label{eq-LV1}
{\cal L}_{\text{V}} = -\frac14F_{\mu\nu}F^{\mu\nu}+g h_{\mu\nu}\tilde{T}_{\text{V}}^{\mu\nu}-\frac12g^2(h_{\mu\rho}h^{\mu\sigma}\partial^\rho A^\nu\partial_\sigma A_\nu-h_{\mu\rho}h^{\nu\sigma}\partial^\rho A_\nu\partial_\sigma A^\mu),
\end{equation}
where $\tilde{T}_{\text{V}}^{\mu\nu}$ is expressed as
\begin{equation}
\tilde{T}_{\text{V}}^{\mu\nu} = -F^{\mu\rho}\partial^\nu A_\rho.
\end{equation}
If $h_{\mu\nu}$ is gauged to be symmetric and traceless, the interaction $g h_{\mu\nu}\tilde{T}_{\text{V}}^{\mu\nu}$ can be changed into $\frac12 g h_{\mu\nu}(T_{\text{V}}^{\mu\nu}+T_{\text{V}}^{\nu\mu})$, where
$T_{\text{V}}^{\mu\nu}$ is the energy-momentum tensor of the photon field, expressed as
\begin{equation}
T_{\text{V}}^{\mu\nu} = -F^{\mu\rho}\partial^\nu A_\rho + \frac14 \eta^{\mu\nu} F_{\rho\sigma}F^{\rho\sigma}.
\end{equation}
The $\eta^{\mu\nu}$ term in $T_{\text{V}}^{\mu\nu}$ has no contribution to the interaction $\frac12 g h_{\mu\nu}(T_{\text{V}}^{\mu\nu}+T_{\text{V}}^{\nu\mu})$ because $h_{\mu\nu}$ is traceless. As in the fermion field case, with the symmetric tensor current, the corresponding Lagrangian is locally invariant only under the symmetric and traceless translation. 

With the replacement of the derivative $\partial_\mu$ with $D_\mu$, the locally $U(1)$ invariant Lagrangian ${\cal L}_{\text{V}}^0$ is transformed into a translation invariant one ${\cal L}_{\text{V}}$. 
As a result, the Lagrangian of Eq.~(\ref{eq-LV1}) is no longer $U(1)$ invariant. This is acceptable because the gravitational interaction is translation invariant rather than $U(1)$ invariant.
By replacing $F_{\mu\nu}$ with $F_{\mu\nu}+gh_{\mu\rho}F^\rho_\nu-gh_{\nu\rho}F^\rho_\mu$, we can obtain the $U(1)$ invariant Lagrangian
\begin{eqnarray}\label{eq-LV2}
\mathcal{L}_{\text{V}}&=&-\frac14(F_{\mu\nu}+gh_{\mu\rho}F^\rho_\nu-gh_{\nu\rho}F^\rho_\mu)(F^{\mu\nu}+gh^{\mu\sigma}F_\sigma^\nu-gh^{\nu\sigma}F_\sigma^\mu) \nonumber \\
&=&-\frac14F_{\mu\nu}F^{\mu\nu}+gh_{\mu\nu}T_{\text{EM}}^{\mu\nu}-\frac12g^2(h_{\mu\rho}h^{\mu\sigma}F^{\rho\nu}F_{\sigma\nu}-h_{\nu\rho}h^{\mu\sigma}F^\rho_\mu F_\sigma^\nu),
\end{eqnarray}
where
$T_{\text{EM}}^{\mu\nu}$ is the symmetric Belinfante-Rosenfeld tensor of the electromagnetic field, expressed as
\begin{equation}
T_{\text{EM}}^{\mu\nu}=F^{\mu\rho}F_\rho^\nu+\frac14\eta^{\mu\nu}F_{\rho\sigma}F^{\rho\sigma}=T_{\text{V}}^{\mu\nu}+\partial_\rho (F^{\mu\rho}A^\nu).
\end{equation}
The difference between the symmetric Belinfante-Rosenfeld tensor $T_{\text{EM}}^{\mu\nu}$ and the asymmetric energy-momentum tensor $T_{\text{V}}^{\mu\nu}$, which corresponds to the conserved tensor directly obtained from N\"{o}ther theory, is the total derivative term $\partial_\rho (F^{\mu\rho}A^\nu)$. We should mention that though the interaction $gh_{\mu\nu}T_{\text{EM}}^{\mu\nu}$ is recognized as the interaction between the photon and gravitational fields, the Lagrangian of Eq.~(\ref{eq-LV2}) is not locally translation invariant. In fact, in the fermion field case, the situation is the same. The spin-$\frac12$ Lagrangian with the gravitational field cannot be locally $U(1)$ gauge invariant. For example, with minimal substitution, if we change the Lagrangian $\mathcal{L}_{\text{F}}$ to be
\begin{equation}
\mathcal{L}_{\text{F}}'= \bar{\psi}\gamma^\mu(\eta_{\mu\nu}+gh_{\mu\nu})(\partial^\nu+ieA^\nu) \psi - m\bar{\psi}\psi,
\end{equation}
the Lagrangian $\mathcal{L}_{\text{F}}'$ will be locally $U(1)$ invariant. However it will no longer be locally translation invariant. For the gravitational interaction, we should have translation invariance instead of $U(1)$ invariance. In this sense, the Lagrangian Eq.~(\ref{eq-LV1}) is more reasonable than that of Eq.~(\ref{eq-LV2}). This is different from the standard model, where the total Lagrangian is invariant under the local $SU(3)_C\times SU(2)_L \times U(1)_Y$ transformation. The QCD Lagrangian is $SU(3)_C$ invariant, and the electroweak Lagrangian is $SU(2)_L\times U(1)_Y$ invariant. Because the currents in the standard model interactions have no derivatives, their sum is both $SU(3)_C$ and $SU(2)_L\times U(1)_Y$ invariant. However,
if gravity is included, we cannot make the total Lagrangian invariant under the $T(4) \times SU(3)_C\times SU(2)_L \times U(1)_Y$ transformation. The tensor current for the gravitational interaction includes the derivative. As a result, the total Lagrangian including the gravitational field will destroy the internal local gauge symmetry, although the other parts without gravitational field ars still invariant under the internal gauge transformations.

\subsection{Lagrangian for gravitational field}
Now, we discuss the Lagrangian for the spin-2 gravitational field. For the spin-2 field $h_{\mu\nu}$, its free Lagrangian is \cite{Adrian}
\begin{equation}
\mathcal{L}^0_{\text{G}}=\frac12\partial_\mu h_{\rho\sigma}\partial^\mu h^{\rho\sigma}-\partial_\mu h^{\mu\nu}\partial^\rho h
_{\rho\nu}+\partial_\mu h\partial_\nu h^{\mu\nu}-\frac12\partial_\mu h\partial^\mu h,
\end{equation}
where $h$ is the trace of the field $h_{\mu\nu}$. The above Lagrangian is the same as the linear approximation of general relativity, where $h_{\mu\nu}$ is a weak field \cite{mtw}. The Lagrangian $\mathcal{L}^0_{\text{G}}$ is only invariant under the infinitesimal local translation. To make the Lagrangian invariant under the general local translation, we must change the derivative $\partial_\mu$ to $D_\mu$. For example, for $\partial_\mu h_{\nu\rho}(x)$, we must change it to be $D_\mu D_\nu V_\rho(x)$, where $V_\rho(x)$ is an arbitrary vector field. This ``background'' vector is introduced together with the derivative because of the transformation property of the gravitational field $h_{\mu\nu}$ Eq.~(\ref{eq-transh}). Under the local translation $D_\mu D_\nu V_\rho(x)$ transforms as
\begin{eqnarray}
D_\mu D_\nu V_\rho(x) &\rightarrow& \left\{\partial_\mu + g\left[h_{\mu\sigma}(x') - \frac1g \partial_\mu \theta_\sigma(x)\right]\partial'^\sigma\right\} \left\{\left[\partial_\nu + g\left(h_{\nu\tau}(x') - \frac1g \partial_\nu \theta_\tau(x)\right)\partial'^\tau \right]V_\rho(x')\right\} \nonumber \\
&=& \left\{\partial_\mu + g\left[h_{\mu\sigma}(x') - \frac1g \partial_\mu \theta_\sigma(x)\right]\partial'^\sigma\right\} \left\{\left [\partial'_\nu + gh_{\nu\tau}(x')\partial'^\tau \right ]V_\rho(x')\right\}
\nonumber \\
&=& \left\{\partial'_\mu + gh_{\mu\sigma}(x')\partial'^\sigma\right\} \left\{\left [\partial'_\nu + gh_{\nu\tau}(x')\partial'^\tau \right ]V_\rho(x')\right\}\nonumber \\
&=& D'_\mu D'_\nu V_\rho(x').
\end{eqnarray}
Therefore, the Lagrangian for the gravitational field can be constructed by $D_\mu D_\nu V_\rho(x)$. The simplest choice is to choose $V_\rho(x) = \frac1g x_\rho$. $\partial_\nu V_\rho(x)$ turns into $\frac1g \mathfrak{N}_{\nu\rho}(x)$, where $\mathfrak{N}_{\nu\rho}(x)$ transforms as
\begin{equation}
\mathfrak{N}_{\nu\rho}(x) \rightarrow \mathfrak{N}_{\nu\rho}(x') + \partial_\nu \theta_\rho(x). 
\end{equation}
In Minkowski spacetime, $\mathfrak{N}_{\nu\rho}(x) =  \mathfrak{N}_{\nu\rho}(x') =  \eta_{\nu\rho}$. In this case, $\frac 1g D_\mu D_\nu x_\rho(x)=D_\mu h_{\nu\rho}(x)$.
Consequently, the locally translation invariant Lagrangian can then be written as
\begin{eqnarray}
\mathcal{L}_{\text{G}}&=&\frac12D_\mu h_{\rho\sigma}D^\mu h^{\rho\sigma}-D_\mu h^{\mu\nu}D^\rho h
_{\rho\nu}+D_\mu hD_\nu h^{\mu\nu}-\frac12D_\mu hD^\mu h\nonumber \\
&=&\mathcal{L}_{\text{G}}^0+gh_{\mu\nu}\tilde{T}_{\text{G}}^{\mu\nu} 
+\frac{g^2}{2}\big(h_{\alpha\mu}h^{\alpha\nu}\partial^\mu h^{\rho\sigma}\partial_\nu h_{\rho\sigma}-h_{\alpha\mu}h^{\alpha\nu}\partial^\mu h\partial_\nu h \nonumber \\
&&-2h^{\mu\rho}h_{\nu\sigma}\partial_\rho h_{\mu\alpha}\partial^\sigma h^{\nu\alpha}+2h_{\mu\rho}h_{\nu\sigma}\partial^\rho h^{\mu\nu}\partial^\sigma h\big),
\end{eqnarray}
where $\tilde{T}_{\text{G}}^{\mu\nu}$ is expressed as
\begin{equation}
\tilde{T}_{\text{G}}^{\mu\nu}=\partial^\mu h_{\rho\sigma}\partial^\nu h^{\rho\sigma}-\partial^\mu h\partial^\nu h-2\partial^\nu h^{\mu\rho}\partial^\sigma h_{\sigma\rho}+\partial^\nu h^{\mu\rho}\partial_\rho h+\partial^\mu h\partial_\rho h^{\rho\nu}.
\end{equation}
With the requirement of translation invariance, the self-interactions of the gravitational field can be obtained, which include the leading order interaction $gh_{\mu\nu}\tilde{T}_{\text{G}}^{\mu\nu}$ and higher order terms proportional to $g^2$.

We obtained the interactions between the gravitational field and other matter fields and its self-interactions based on the translation invariance. In the next section, we study the bending of light when it passes through the sun as an example. The total related Lagrangian can be written as
\begin{eqnarray}
\mathcal{L}_{\text{TOT}} &=& \bar{\psi}(x)\gamma^\mu(\partial_\mu + ie A_\mu(x))\psi(x) +gh_{\mu\nu}(x) \tilde{T}_{\text{F}}^{\mu\nu} - m\bar{\psi}(x)\psi(x) \nonumber \\
&& -\frac14F_{\mu\nu}F^{\mu\nu}+g h_{\mu\nu}\tilde{T}_{\text{V}}^{\mu\nu}-\frac{g^2}{2}(h_{\mu\rho}h_{\mu\sigma}\partial^\rho A^\nu\partial^\sigma A^\nu-h_{\mu\rho}h_{\nu\sigma}\partial^\rho A^\nu\partial^\sigma A^\mu) \nonumber \\
&&+ \frac12\partial_\mu h_{\rho\sigma}\partial^\mu h^{\rho\sigma}-\partial_\mu h^{\mu\nu}\partial^\rho h
_{\rho\nu}+\partial_\mu h\partial_\nu h^{\mu\nu}-\frac12\partial_\mu h\partial^\mu h+gh_{\mu\nu}\tilde{T}_{\text{G}}^{\mu\nu} \nonumber \\
&&+\frac{g^2}{2}\big(h_{\alpha\mu}h^{\alpha\nu}\partial^\mu h^{\rho\sigma}\partial_\nu h_{\rho\sigma}-h_{\alpha\mu}h^{\alpha\nu}\partial^\mu h\partial_\nu h-2h^{\mu\rho}h_{\nu\sigma}\partial_\rho h_{\mu\alpha}\partial^\sigma h^{\nu\alpha} \nonumber \\
&&+2h_{\mu\rho}h_{\nu\sigma}\partial^\rho h^{\mu\nu}\partial^\sigma h\big),
\end{eqnarray}
where the other gauge fields for the strong and weak interactions are not included. They can be easily added to the above Lagrangian. With the Lagrangian, the equation of motion for the gravitational field can be obtained as
\begin{equation}\label{eq-EOMh}
H_0^{\mu\nu}+gH_1^{\mu\nu}+g^2H_2^{\mu\nu} = g\left(\tilde{T}_{\text{F}}^{\mu\nu} + \tilde{T}_{\text{V}}^{\mu\nu}\right)-g^2(h^{\mu\rho}F_{\rho\sigma}F^{\nu\sigma}+h_{\rho\sigma}F^{\mu\rho}F^{\nu\sigma}),
\end{equation}
where
\begin{align}
H_0^{\mu\nu}&=\square h^{\mu\nu}-\eta^{\mu\nu}\square h-\partial^\mu\partial_\rho h^{\rho\nu}-\partial^\nu\partial_\rho h^{\rho\mu}+\partial^\mu\partial^\nu h+\eta^{\mu\nu}\partial_\rho\partial_\sigma h^{\rho\sigma},\\
H_1^{\mu\nu}&=2\partial_\rho h^{\rho\sigma}\partial_\sigma h^{\mu\nu}+2h^{\rho\sigma}\partial_\rho\partial_\sigma h^{\mu\nu}-2\eta^{\mu\nu}\partial_\rho h^{\rho\sigma}\partial_\sigma h-2\eta^{\mu\nu}h^{\rho\sigma}\partial_\rho\partial_\sigma h\notag\\
&\phantom{=}-2\partial_\rho h^{\rho\mu}\partial_\sigma h^{\sigma\nu}-2h^{\rho\mu}\partial_\rho\partial_\sigma h^{\sigma\nu}-2\partial^\mu h_{\rho\sigma}\partial^\rho h^{\sigma\nu}-2h_{\rho\sigma}\partial^\mu\partial^\rho h^{\sigma\nu}\notag\\
&\phantom{=}+\partial_\rho h^{\rho\mu}\partial^\nu h+h^{\rho\mu}\partial_\rho\partial^\nu h+\eta^{\mu\nu}\partial_\rho h_{\alpha\beta}\partial^\beta h^{\alpha\rho}+\eta^{\mu\nu}h_{\alpha\beta}\partial_\rho \partial^\beta h^{\alpha\rho}\notag\\
&\phantom{=}+\partial^\mu h^{\rho\nu}\partial_\rho h+h^{\rho\nu}\partial^\mu\partial_\rho h+\eta^{\mu\nu}\partial_\rho h^{\rho\sigma}\partial^\alpha h_{\alpha\sigma}+\eta^{\mu\nu}h^{\rho\sigma}\partial_\rho \partial^\alpha h_{\alpha\sigma}-\tilde{T}_{\text{G}}^{\mu\nu},\\
H_2^{\mu\nu}&=h^{\alpha\sigma}\partial^\rho h_{\alpha\rho}\partial_\sigma(h^{\mu\nu}-\eta^{\mu\nu}h)+h_{\alpha\rho}\partial^\rho h^{\alpha\sigma}\partial_\sigma(h^{\mu\nu}-\eta^{\mu\nu}h)+h_{\alpha\rho}h^{\alpha\sigma}\partial^\rho\partial_\sigma(h^{\mu\nu}-\eta^{\mu\nu}h)\notag\\
&\phantom{=}-2\partial_\rho(h^{\rho\mu}h_{\sigma\alpha}\partial^\sigma h^{\alpha\nu})+\partial_\rho(h^{\rho\mu}h^{\sigma\nu}\partial_\sigma h)+\eta^{\mu\nu}\partial^\rho(h_{\rho\alpha}h_{\sigma\beta}\partial^\sigma h^{\alpha\beta})-h^{\mu\rho}\partial^\nu h^{\alpha\beta}\partial_\rho h_{\alpha\beta}\notag\\
&\phantom{=}+h^{\mu\rho}\partial^\nu h\partial_\rho h+2h^{\alpha\sigma}\partial^\nu h^{\mu\beta}\partial_\sigma h_{\alpha\beta}-h_{\rho\sigma}\partial^\nu h^{\mu\rho}\partial^\sigma h-h_{\rho\sigma}\partial^\rho h^{\sigma\nu}\partial^\mu h.
\end{align}
With the same gauge as Eq.~(\ref{eq-gauge}), the above equation can be simplified to
\begin{eqnarray}
&& \square h^{\mu\nu}+g(2h^{\rho\sigma}\partial_\rho\partial_\sigma h^{\mu\nu}-2\partial^\mu h_{\rho\sigma}\partial^\rho h^{\sigma\nu}-2h_{\rho\sigma}\partial^\mu\partial^\rho h^{\sigma\nu}+\eta^{\mu\nu}\partial_\rho h_{\alpha\beta}\partial^\beta h^{\alpha\rho}-\partial^\mu h^{\rho\sigma}\partial^\nu h_{\rho\sigma}) \nonumber \\
&& +g^2\left(h_{\alpha\rho}\partial^\rho h^{\alpha\sigma}\partial_\sigma h^{\mu\nu}+h_{\alpha\rho}h^{\alpha\sigma}\partial^\rho\partial_\sigma h^{\mu\nu}-2h^{\rho\mu}\partial_\rho(h_{\sigma\alpha}\partial^\sigma h^{\alpha\nu})+\eta^{\mu\nu}h_{\rho\alpha}\partial^\rho(h_{\sigma\beta}\partial^\sigma h^{\alpha\beta}) \right . \nonumber \\
&& ~~~~~~\left.-h^{\mu\rho}\partial^\nu h^{\alpha\beta}\partial_\rho h_{\alpha\beta}+2h^{\alpha\sigma}\partial^\nu h^{\mu\beta}\partial_\sigma h_{\alpha\beta}
+ h^{\mu\rho}F_{\rho\sigma}F^{\nu\sigma}+h_{\rho\sigma}F^{\mu\rho}F^{\nu\sigma} \right) \nonumber \\ 
&& = g\left(\tilde{T}_{\text{F}}^{\mu\nu} + \tilde{T}_{\text{V}}^{\mu\nu}\right).
\end{eqnarray}
On the right hand side of the equation, the tensor currents are for all matter fields that generate the gravitational field. To study the bending of light by the sun, the energy-momentum tensor of the sun is dominant. 

\section{Bending of light}\label{sec-3}
We now discuss the bending of light caused by the sun in Minkowski spacetime with the Lagrangian obtained in the previous section. We first obtain the solution of the gravitational field generated by the sun. It is difficult to solve the full differential equation (\ref{eq-EOMh}); therefore, we solve the equation perturbatively. At leading order, where the self-interactions of the gravitational field are neglected, the equation for leading order $h_{(0)}^{\mu\nu}$ is expressed as
\begin{equation}\label{eq-leading}
\square h_{(0)}^{\mu\nu}-\eta^{\mu\nu}\square h_{(0)}-\partial^\mu\partial_\rho h_{(0)}^{\rho\nu}-\partial^\nu\partial_\rho h_{(0)}^{\rho\mu}+\partial^\mu\partial^\nu h_{(0)}+\eta^{\mu\nu}\partial_\rho\partial_\sigma h_{(0)}^{\rho\sigma} =g T_{\text{sun}}^{\mu\nu},
\end{equation}
where $T_{\text{sun}}^{\mu\nu}$ is the tensor current for the classical object sun. For a microscopic particle, the tensor $\tilde{T}^{\mu\nu}$ is different from the symmetric energy-momentum tensor $T^{\mu\nu}$. 
The difference between them is a total differential and a term proportional to the free Lagrangian ${\cal L}^0$. For example, for Dirac field, $\tilde{T}_{\text{F}}^{\mu\nu}-T_{\text{F}}^{\mu\nu} =-\frac{i}{16}\partial_\rho\left(\bar{\psi}\{\gamma^\mu,[\gamma^\nu,\gamma^\rho]\}\psi\right) +\eta^{\mu\nu}\mathcal{L}^0_{\text{F}}$, and for the scalar field, 
$\tilde{T}_{\text{S}}^{\mu\nu}-T_{\text{S}}^{\mu\nu} = \eta^{\mu\nu}{\cal L}^0_{\text{S}}$. When studying gravity for an on-shell (${\cal L}^0 = 0$) classical object, one can express the tensor current of the object as \cite{mtw}
\begin{equation}
T_{\text{cl}}^{\mu\nu}(x)=\int m\frac{d z^\mu}{d\tau}\frac{d z^\nu}{d\tau}\delta^4(\mathbf{x}-\mathbf{z}(\tau))d\tau
\end{equation}
without specifying its spin. Here, $z^\mu(\tau)$ represents the particle's world line, and $m$ is the mass of the object. In particular, the classical tensor for a static star $T_{\text{sun}}^{\mu\nu}$ has only one non-zero component expressed as
\begin{equation}\label{eq-emofsun}
T_{\text{sun}}^{00}(x)=M\delta^{(3)}(\mathbf{x}),
\end{equation}
where $M$ is the mass of the sun. Eq. (\ref{eq-leading}) is comparable to the metric $\mathfrak{h}_{\mu\nu}$ in the linear approximation of general relativity, where
$\mathfrak{h}_{\mu\nu}=g_{\mu\nu} - \eta_{\mu\nu}$, and $g_{\mu\nu}$ is the metric tensor of curved spacetime. In our case, the gravitational field $h_{\mu\nu}$ is a dimensional quantity related to $\mathfrak{h}_{\mu\nu}$ via $gh_{\mu\nu}=\mathfrak{h}_{\mu\nu}$. The coupling constant is related to Newton's constant $G$ via $g^2=16\pi G$. The value of $g$ is an order of magnitude similar to the Planck mass's reciprocal. This means that the strength of the $h_{\mu\nu}$ coupling to itself and other fields is weak. 

We should note that when the high order terms are neglected, the equation for the gravitational field at leading order is no longer translation invariant. The gauge condition of Eq.~(\ref{eq-gauge}) cannot be completely satisfied. Instead, we can choose the gauge condition as
\begin{equation}\label{eq-gauge1}
h_{\mu\nu}=h_{\nu\mu}, \quad \partial^\mu h_{\mu\nu}=0.
\end{equation}
Then, the equations for the non-zero components of $h_{(0)}^{\mu\nu}$ at leading order become
\begin{eqnarray}\label{eq-leading1}
-\nabla^2 h_{(0)}^{00}+\nabla^2 h_{(0)}&=&4\sqrt{\pi G} M\delta^{(3)}(\bold{x}), \\
-\nabla^2 h_{(0)}^{ij}-\nabla^2 h_{(0)}\delta^{ij}+
\partial^i\partial^j h_{(0)}&=&0.
\end{eqnarray}
The solutions for the above equations are
\begin{equation}\label{eq-h-gauge1}
h_{(0)}^{00}=\frac12\sqrt{\frac{G}{\pi}}\frac{M}{r},\quad  
h_{(0)}^{ij}=\frac14\sqrt{\frac{G}{\pi}}\left(\frac{M}{r}\delta^{ij}+\frac{Mx^ix^j}{r^3}\right).
\end{equation}
It is easy to check that the solutions satisfy the condition $\partial_\mu h_{(0)}^{\mu\nu}=0$.

We can also choose another type of gauge condition widely used in solving the metric $\mathfrak{h}_{\mu\nu}$ in the linear approximation of general relativity \cite{mtw,Strau2}. The condition is
\begin{equation}\label{eq-gauge'}
h_{\mu\nu}=h_{\nu\mu}, \quad 2\partial_\mu h^{\mu\nu}=\partial^\nu h,
\end{equation}
which is also weaker than the condition of  Eq.~(\ref{eq-gauge}). It is not required that $h=0$ and $\partial_\mu h^{\mu\nu}=0$, but there is a relationship between them. With the above gauge condition, at leading order, the equations for the non-zero component of $h_{\mu\nu}$ turn into
\begin{eqnarray}\label{eq-leading2}
-\nabla^2 \left(h_{(0)}^{00}-\frac12 h_{(0)}\right)&=&4\sqrt{\pi G}M\delta^{(3)}(\bold{x}), \\
-\nabla^2 \left(h_{(0)}^{ii}+\frac12 h_{(0)}\right)&=&0.
\end{eqnarray}
The corresponding solutions are
\begin{equation}\label{eq-h-gauge2}
h_{(0)}^{00}=\frac12\sqrt{\frac{G}{\pi}}\frac{M}{r},\quad 
h_{(0)}^{ii}=\frac12\sqrt{\frac{G}{\pi}}\frac{M}{r}.
\end{equation}
With the obtained gravitational field $h_{\mu\nu}$, we can now calculate the deflection angle of light when it passes through the sun.

The action of a classical particle in the presence of a gravitational field is written as 
\begin{equation}
I_{\text{particle}}=-m\int(\eta_{\mu\nu}+gh_{\mu\nu})\frac{d z^\mu}{d\tau}\frac{d z^\nu}{d\tau}{d\tau}.
\end{equation}
With the variational principle, the equation of motion for the particle is expressed as
\begin{equation}\label{eq-EOM}
\frac{d^2 z^\mu}{d\tau^2}+\Gamma^\mu_{\rho\sigma}\frac{d z^\rho}{d\tau}\frac{d z^\sigma}{d\tau}=0,
\end{equation}
where
\begin{equation}
\Gamma^\mu_{\rho\sigma}\equiv\frac12(\eta^{\mu\nu}+gh^{\mu\nu})(\partial_\rho gh_{\nu\sigma}+\partial_\sigma gh_{\nu\rho}-\partial_\nu gh_{\rho\sigma}).
\end{equation}
This equation is similar to the geodesic equation in the general relativity case. The above equation is not mass dependent and is suitable for a particle with any mass, including photons. With the definition $P_\mu\equiv(\eta_{\mu\nu}+gh_{\mu\nu})p^\nu$ where $p^\nu= \frac{dz^\nu}{d\tau}$, Eq.~(\ref{eq-EOM}) yields
\begin{equation}\label{eq-EOMphoton}
dP_\mu=\frac{g}{2}\partial_\mu h_{\rho\sigma}p^\rho dz^\sigma.
\end{equation}
Suppose there is a photon moving in the $x$--$y$ plane ($z=0$). The direction of its initial momentum at $(x=-\infty,y=R,z=0)$ is parallel to $x$ axis. The trajectory of the photon will be deflected by the sun when it passes through the sun. Because the deflection of the photon is slight, we can assume that $p^x$ is a constant and $|p^y|\ll p^0\approx p^x$. 
With the obtained $h_{\mu\nu}$ in Eq.~(\ref{eq-h-gauge1}) for the gauge condition of Eq.~(\ref{eq-gauge1}), the equation of motion \eqref{eq-EOMphoton} for  $\mu=2$ is expressed as
\begin{equation}\label{eq-bend1}
d P_y=-\left(\frac{3GMR}{2\sqrt{x^2+R^2}^3}+\frac{3GMx^2R}{2\sqrt{x^2+R^2}^5}\right)p^x dx.
\end{equation}
After the $x$ integral we have
\begin{equation}
P_y(x=+\infty)=-p^x\int_{-\infty}^{\infty}\left(\frac{3GMR}{2\sqrt{x^2+R^2}^3}+\frac{3GMx^2R}{2\sqrt{x^2+R^2}^5}\right)dx=-\frac{4GM}{R}p^x.
\end{equation}
The deflection angle of light is then obtained as 
\begin{equation}\label{eq-bendresult1}
\Delta\phi=-\frac{p^y(x=+\infty)}{p^x}=-\frac{P_y(x=+\infty)}{p^x}=\frac{4GM}{R}. 
\end{equation}
For the other gauge condition Eq.~(\ref{eq-gauge'}), the corresponding equation of motion for $\mu=2$ is
\begin{equation}\label{eq-bend2}
d P_y=-\frac{2GMR}{\sqrt{x^2+R^2}^3}p^x dx.
\end{equation}
This equation is different from Eq.~(\ref{eq-bend1})
owing to the different $h_{\mu\nu}$ obtained in the different gauge. It is interesting that after the $x$ integral of the above equation, we obtain the same $P_y(x=+\infty)$ as 
\begin{equation}
P_y(x=+\infty)=-p^x\int_{-\infty}^{\infty}\frac{2GMR}{\sqrt{x^2+R^2}^3}dx=-\frac{4GM}{R}p^x.
\end{equation}
With the mass and radius of the sun, we can derive the deflection angle $\Delta\phi=1.75''$.
Therefore, for different gauge conditions, although the corresponding $h_{\mu\nu}$ forms are different, the obtained deflection angle of light is the same. It is also the same as the result in general relativity and experimental observations \cite{will,Shapiro}. 

We now estimate the deflection angle including the self-interaction of gravitational field. For simplicity, we work in the gauge condition of Eq.~(\ref{eq-gauge'}). At next-to-leading order, the equations for the non-zero components of the gravitational field are expressed as
\begin{align}
-\nabla^2\left(h_{(1)}^{00}-\frac12h_{(1)}\right)&=4\sqrt{\pi G}M\delta^{(3)}(\bold{x})-4G\sqrt{\frac{G}{\pi}}\frac{M^2}{r^4},\\
-\nabla^2\left(h_{(1)}^{ij}+\frac12\delta^{ij}h_{(1)}\right)&=2G\sqrt{\frac{G}{\pi}}\frac{M^2}{r^6}x^ix^j.
\end{align}
The solutions for the above equations are
\begin{equation}\label{eq-h-gauge2'}
h_{(1)}^{00}=\frac12\sqrt{\frac{G}{\pi}}\left(\frac{M}{r}+\frac{GM^2}{r^2}\right),\quad h^{ij}_{(1)}=\frac12\sqrt{\frac{G}{\pi}}\left(\frac{M}{r}\delta^{ij}+\frac{2GM^2}{r^2}\delta^{ij}+\frac{GM^2}{r^4}x^ix^j\right).
\end{equation}
With the same method, we obtain
\begin{eqnarray}
P_y(x=+\infty)&=&-p^x\int_{-\infty}^{\infty}\left(\frac{2GMR}{\sqrt{x^2+R^2}^3}+\frac{6G^2M^2R}{\sqrt{x^2+R^2}^4}+\frac{4G^2M^2x^2R}{\sqrt{x^2+R^2}^6}\right)dx \nonumber \\
&=&-\left(\frac{4GM}{R}+\frac{7\pi}{2}\frac{G^2M^2}{R^2}\right)p^x.
\end{eqnarray}
The deflection angle of light is then
\begin{equation}\label{eq-bendresult3}
\Delta\phi=\frac{4GM}{R}+\frac{7\pi}{2}\frac{G^2M^2}{R^2},
\end{equation}
where the second term on the right hand side is the next-to-leading order contribution, which is highly suppressed by the factor $\frac{GM}{R}$. Numerically, the correction to $\Delta\phi$ at next-to-leading order is negligible, at approximately $1.0''\times 10^{-5}$.

For the photon, the gravitational field generated by the sun is very weak and the deflection angle of light is small when it passes through the sun. In this case, the self-interaction of the gravitational field is negligible. Its correction to $\Delta\phi$ is not visible. For both $h_{\mu\nu}$ and $\Delta\phi$, the high order contributions are suppressed by the factor $\frac{GM}{R}$.
For more compatible stars, for example, the white dwarf ZTFJ$190132.9+145808.7$, with a mass and radius of $M\approx1.35M_\odot$ and $R\approx2140$ $\text{km}$ \cite{Caiazzo},
the factor $\frac{GM}{R}$ is $9.3\times10^{-4}$, which is $439$ times larger than that of the sun. As a result, at leading and next-to-leading order, the deflection angles $\Delta\phi$ are $12.79'$ and $12.83'$, respectively. For neutron star HESS J1731-347, with a mass and radius of $M\approx0.77M_\odot$ and $R\approx10.4$ $\text{km}$ \cite{Doroshenko}, the factor $\frac{GM}{R}$ is $0.11$, and the contribution from the self-interaction of gravitational field may be as large as $30\%$ of leading order contribution.

We should mention hat the gauge theories of gravity have been proposed with respect to various external groups, such as the Lorentz group, translation group, and Poincar\'{e} group in the 1960s and 1970s \cite{utiyama1,hayashi0,hayashi1,hayashi2,hayashi3,terg2,utiyama2,kibble}. In particular, a series of papers by Hayashi et al discussed in detail how to construct the gauge theory of gravity via the translation group \cite{hayashi0,hayashi1,hayashi2,hayashi3}. Compared with Ref.~\cite{hayashi0}, although the basic idea of this study is similar, for example, the gravitational field is introduced by changing the global translation symmetry into a local one, there are several major differences. First, our Lagrangian is invariant under the finite translation transformation, whereas the Lagrangian in Ref.~\cite{hayashi0} is invariant under the infinitesimal transformation. As a result, the corresponding transformation of gravitational field $h_{\mu\nu}$ in this study is always associated with the derivative, as shown in Eq.~(\ref{eq-transh}). The transformation of $h_{\mu\nu}$ itself can be obtained order by order as Eqs.~(\ref{eq-trans1}) and (\ref{eq-trans2}). The transformation property of the gravitational field in Ref.~\cite{hayashi0} is the same as our leading order formula if high order terms are neglected. Second, our gravitational field $h_{\mu\nu}$ has nothing to do with the metric. The metric $g_{\mu\nu}$ in our Lagrangian is always $\eta_{\mu\nu}$. The gravitational field $h_{\mu\nu}$ is an independent quantity. Owing to the translation invariance of the Lagrangian, we can choose a different ``gauge" for $h_{\mu\nu}$. Conversely, in Ref.~\cite{hayashi0}, the gravitational field $b_{k}^\mu$ was proved to be a vierbein (tetrad) field and related to the metric. As expressed in the reference, it is necessary to define the field $b_{k\mu}$ inverse to $b_k^\mu$ with $b_{k\mu}b_l^\mu=\delta_{kl}$ and $b$ with $b=\text{det}(b_{k\mu})$. The metric tensor was constructed as $g^{\mu\nu}=\eta^{kl}b_k^\mu b_l^\nu$, and the invariant volume element was $bd^4x$ instead of $d^4x$. The underlying Minkowski spacetime was deformed after localizing the translation, thus one had to reconstruct the emerging geometry \cite{Hehl}. Third, based on the local translation invariance, the interactions between the $h_{\mu\nu}$ and matter fields with spin 0, $\frac12$ and 1 are obtained. Except the interaction between the $h_{\mu\nu}$ and spin-$\frac12$ field, there are high-order interactions. In particular, the interaction between $h_{\mu\nu}$ and the electromagnetic field $A_\mu$ is not locally $U(1)$ invariant. In addition, our Lagrangian for the gravitational field is obtained from the free Lagrangian for the spin-2 field with the requirement of locally translation invariance. The obtained Lagrangian is also different from that of Ref.~\cite{hayashi0}. Finally, we describe gravity in the same frame as that for the other interactions in the standard model. The result obtained with our Lagrangian is different from that with general relativity. While previous gauge theories of gravity lead to the so called ``new general relativity", which is the teleparallel equivalent of Einstein's general relativity \cite{hayashi3}, the ``vierbein" approach can be regarded as another formalism to derive Einstein's equation \cite{modify,Santos1}.

\section{summary}\label{sec-4}
Based on the requirement that the total Lagrangian should be locally translation invariant, we obtain the interactions between the gravitational field and matter fields in Minkowski spacetime. For the spin-$\frac12$ field, the tensor current, which couples to the gravitational field $h_{\mu\nu}$, is not symmetric. This differs from the well known symmetric Belinfante--Rosenfeld energy--momentum tensor of the Dirac field. Only when the symmetric and traceless gauge is chosen will the gravitational field couple with the conserved symmetric tensor. With this gauge, the Lagrangian is no longer invariant under the general translation group but is only invariant under the symmetric and traceless translation. This is also true for the spin-1 case, where the current couples to the gravitational field, which is not a symmetric conserved energy--momentum tensor. In addition, forthe  photon field, the inclusion of the gravitational interaction will destroy the local $U(1)$ invariance. For the spin-0 case, although the tensor coupled to $h_{\mu\nu}$ is symmetric, it is still different from the conserved energy-momentum tensor because of the $\eta_{\mu\nu}$ term. For both the spin-0 and 1 cases, besides the leading order interaction $gh_{\mu\nu} \tilde{T}^{\mu\nu}$, the local translation invariance results in high order interactions proportional to $g^2$. The translation invariance also leads to leading and next-to-leading order self-interactions in the Lagrangian for the spin-2 gravitational field.

We discuss the deflection of light with the interaction between the photon and gravitational field in Minkowski spacetime. For two different choices of gauge condition, the obtained gravitational fields $h_{\mu\nu}$ are different. However, the deflection angles $\Delta \phi$ in the two cases are the same. The obtained angle is also the same as that in general relativity, although the basic scenario is completely different. The contribution from the self-interaction of gravitational field is suppressed by the factor $\frac{GM}{R}$. It is negligible and causes no visible effect on $\Delta\phi$ of light when it passes through the sun. For more compact stars, for example, a neutron star, the self-interaction may contribute as large as $30\%$ of the leading order contribution. Therefore, the difference between the gravity described in our method and that in general relativity is significant when gravity is strong. This may provide a new scenario for our Universe. It is also interesting that gravity can be described in a similar way to the interactions in the standard model, which is based on local symmetry in Minkowski spacetime.

\section*{Acknowledgments}

This work is supported by the National Nature Science Foundation of China (NSFC) under Grant No. 11975241.

\end{document}